\documentclass{ws-ijmpd}
\usepackage[super,compress]{cite}
\begin{document}

\markboth{S. D. Maharaj, D. Kileba Matondo, P. Mafa Takisa}
{}

%
\catchline{}{}{}{}{}
%

\title{A family of Finch and Skea relativistic stars}

\author{S. D. Maharaj\footnote{maharaj@ukzn.ac.za}, D. Kileba Matondo\footnote{dkilebamat@gmail.com} \, and  P. Mafa Takisa\footnote{pmafatakisa@gmail.com} }

\address{Astrophysics and Cosmology Research Unit,\\ School of Mathematics, Statistics and Computer Science,\\ University of KwaZulu-Natal, 
\\Private Bag X54001, Durban 4000, \\
South Africa. 
}

\maketitle

\begin{history}
\received{Day Month Year}
\revised{Day Month Year}
\end{history}

\begin{abstract}
Several new families of exact solution to the Einstein-Maxwell system of differential equations are found for anisotropic charged matter. The spacetime geometry is that of Finch and Skea which satisfies all criteria for physical acceptability. The exact solutions can be expressed in terms of elementary functions, Bessel functions and modified Bessel functions. When a parameter is restricted to be an integer then the special functions reduce to simple elementary functions. The uncharged model of Finch and Skea (\emph{Class. Quantum Grav.} \textbf{6} (1989), 467) and the charged model of Hansraj and Maharaj (\emph{Int. J. Mod. Phys. D} \textbf{8} (2006), 1311) are regained as special cases. The solutions found admit a barotropic equation of state. A graphical analysis indicates that the matter and electric quantities are well behaved.
\end{abstract}




\section{Introduction}
The Einstein-Maxwell system of equations has generated much interest recently. Exact solutions may be used to describe the dynamics of charged anisotropic matter in a relativistic stellar setting. The modelling of highly compact objects such as dark energy stars, gravastars, ultradense stars and neutron stars then becomes possible in general relativity. Stars with anisotropic pressures and an electric field have been studied by Maurya and Gupta \cite{hh1}, Maurya \textit{et al.} \cite{hh2}, Pandya \textit{et al.} \cite{hh3}, Bhar \textit{et al.} \cite{hh4}, Fatema and Murad \cite{hh5}, Murad and Fatema \cite{hh6} and Murad \cite{hh7}. Solutions with an equation of state may be related to observed astronomical objects as shown by Maharaj and Mafa Takisa \cite{hh8}, Mafa Takisa \textit{et al.} \cite{hh9, hh10} and Sunzu \textit{et al.} \cite{hh11, hh12}.

In spite of numerous exact solutions that have been found with a static spherically symmetric field only a few families of models are known which satisfy all the criteria for a physically acceptable relativistic star. An ansatz that does lead to a physically valid model is that of Finch and Skea \cite{hh13} with uncharged matter. Charged Finch-Skea stars were found by Hansraj and Maharaj \cite{hh14}; these models are given in terms of Bessel functions and obey a barotropic equation of state. Tikekar and Jotania \cite{hh15} found a two parameter family of solutions describing strange stars and other compact distributions of matter in equilibrium. Stars with a quadratic equation of state with the Finch-Skea geometry were analysed by Sharma and Ratanpal \cite{hh16}. This category of stars was extended by Pandya \textit{et al.} \cite{hh17} for a generalized form of the gravitational potential. Kalam \textit{et al.} \cite{hh18} proposed quintessence stars with both dark energy and anisotropic pressures. Strange stars admitting a Chaphygin equation of state were investigated by Bhar \cite{hh19}. The Finch and Skea \cite{hh13} geometry has been studied in matter distributions with lower and higher dimensions. Banerjee \textit{et al.} \cite{hh20} produced a class of interior solutions, corresponding to the BTZ exterior spacetime \cite{hh21}, in $2+1$ dimensions. Bhar \textit{et al.} \cite{hh22} also produced anisotropic stars in $2+1$ dimensions and a quark equation of state. In higher dimensions the Finch-Skea metrics, and generalisations, also arise as shown by Patel \textit{et al.} \cite{hh23} and Chilambwe and Hansraj \cite{hh24}. It is interesting to observe that the Finch-Skea spacetimes also arise in the $5$-dimensional Einstein-Gauss-Bonnet modified theory of gravity in Hansraj \textit{et al.} \cite{hh25}. This suggests that the Finch-Skea geometry may play an important role in more general Lovelock polynomials with a Lagrangian containing higher order terms.

The above references highlight the importance of the Finch and Skea \cite{hh13} potentials in many different physical applications. We therefore perform a systematic study of the Einstein-Maxwell equations with the Finch-Skea geometry in the presence of anisotropy and charge as it satisfies all physical requirements for a general relativistic stellar configuration and is widely used in the modelling process. Hansraj and Maharaj \cite{hh14} found the charged analogue of the Finch-Skea star. In this investigation we extend the Hansraj and Maharaj approach by adding anisotropy to the field equations. We generate the master gravitational equation in Sect.~\ref{sec2} which is obtained with the help of the Einstein-Maxwell system. We make a particular choice for one of the gravitational potentials, the electric field intensity and the anisotropic term.
 Three classes of solution are possible depending on the 
solution of the differential equation  $4(1+ax)\ddot{y}-2a\dot{y}+(a^2-\alpha)y=0$ and the value of the 
quantity $a^{2}-\alpha$. 
In Sect.~\ref{sec3} we treat the case where $a^{2}-\alpha = 0$. In Sect.~\ref{sec4} we consider the case $a^{2}-\alpha > 0$ and we set $a=-1, 1, 3$. For these values of $a$ we find new classes of exact solution to the Einstein-Maxwell system in terms of elementary functions. In Sect.~\ref{sec5} our study concerns the case $a^{2}-\alpha < 0$. As in the previous section we make the choices $a=-1, 1, 3$ and new classes of exact solutions to the Einstein-Maxwell system are obtained in terms of elementary functions. The equation of state is established in Sect.~\ref{sec6} for a particular model. The other classes of models also admit an equation of state. The physical analysis of the charged anisotropic model is presented in Sect.~\ref{sec7} with graphs generated for particular parameter values for the electric field. Concluding remarks are made in Sect.~\ref{sec8}.

\section{The model\label{sec2}}
The line element has the form
\begin{equation}
\label{M1}
ds^2 = -e^{2\nu(r)}dt^{2}+e^{2\lambda(r)}dr^{2}+r^{2}(d\theta^{2}+\sin^{2}\theta d\phi^{2}),
\end{equation}
where $\nu(r)$ and $\lambda(r)$ are the potentials for a static spherical field. We now introduce the transformation 
\begin{eqnarray}
\label{M2}
x = Cr^{2},~~ Z(x) = e^{-2\lambda (r)},~~ A^{2}y^{2}(x) = e^{2\nu(r)},
\end{eqnarray}
where $A$ and $C$ are constants. This transformation was first used by Durgapal and Bannerji \cite{hh26}. The line element ($\ref{M1}$) then has the form
\begin{eqnarray}
\label{M3}
ds^{2}&=&-A^{2}y^{2}(x)dt^{2}+\frac{1}{4CxZ(x)}dx^{2} \nonumber\\ &&
+ \frac{x}{C}\left(d\theta^{2}+\sin^{2}\theta d\phi^{2}\right),
\end{eqnarray}
in terms of the variable $x$. The Einstein-Maxwell field equations become
\begin{subequations} 
\label{M4}
\begin{eqnarray}
\label{M4a} 
\frac{\rho}{C}&=& -2\dot{Z}+\frac{1-Z}{x}-\frac{\ E^2}{2C},\\
\label{M4b} 
\frac{p_{r}}{C}&=&4Z\frac{\dot{y} }{y}+\frac{Z-1}{x}+\frac{\ E^2}{2C},\\
\label{M4c} 
\frac{p_{t}}{C}&=&4xZ\frac{\ddot{y}}{y}+(4Z+2x\dot{Z})\frac{\dot{y} }{y}+\dot{Z}-\frac{\ E^2}{2C},\\
\label{M4d}
\frac{\sigma^2}{C}&=&\frac{4Z}{x}(x\dot{E}+E)^2,
\end{eqnarray}
\end{subequations}
in terms of the new variables. In the above $\rho$ is the energy density, $p_{r}$ is the radial pressure, $p_{t}$ is the tangential pressure, $E$ is the electric field and $\sigma$ is the charge density. The conservation equation is
\begin{eqnarray}
\label{M5}
\frac{dp_{r}}{dx}&=&-\frac{1}{x}\left[p_{r}-p_{t}+x(\rho+p_{r})\frac{d\nu}{dx}\right]+\frac{E}{x}\frac{d}{dx}(xE).
\end{eqnarray}
The mass of the graviting object contained within a radius $x$ of the sphere is
\begin{equation}
\label{M6}
M(x)=\frac{1}{4C^{3/2}}\int^{x}_{0}\sqrt{\omega}\rho(\omega)\ d\omega.
\end{equation}
This quantity is sometimes called the mass function and is important for comparison with observations.

For a physically realistic relativistic star the equation of state is complex and  depends
 on parameters such as the temperature, the number  fraction of a specific particle interior species and strong entropy gradients. As
 a simplifying assumption for a charged anisotropic matter distribution we assume the barotropic relationship
\begin{equation}
\label{M7}
p_{r} = p_{r}(\rho).
\end{equation}
From $(\ref{M4b})$ and $(\ref{M4c})$ we can write 
\begin{equation}
\label{M8}
4xZ\frac{\ddot{y}}{y} + 2x\dot{Z}\frac{\dot{y}}{y} + \dot{Z} - \frac{Z-1}{x} - \frac{E^{2}}{C} = \frac{\Delta}{C},
\end{equation}
where $\Delta = p_{t}-p_{r}$ is the measure of anisotropy. We can solve the Einstein-Maxwell field equations by choosing specific forms for the gravitational potential $Z$, the electric field intensity $E$ and anisotropy $\Delta$ which are physically reasonable. Therefore we make the choices 
\begin{subequations}
\label{M9}
\begin{eqnarray}
\label{M9a}
 Z&=&\frac{1}{1+ax},\\
\label{M9b} 
\frac{E^2}{C}&=&\frac{(\alpha - \beta)x}{(1+ax)^2},\\
\label{M9c}
\frac{\Delta}{C}&=&\frac{\beta x}{(1+ax)^2},
\end{eqnarray}
\end{subequations}
where  $a$, $\alpha$, $\beta$ are real constants. The electric field $E$ depends on the real parameters $\alpha$ and $\beta$. The form $(\ref{M9b})$ is physically reasonable since $E^2$ remains regular and positive throughout the sphere if  $\alpha > \beta$. In addition the field intensity $E$ becomes zero at the stellar centre and attains a maximum value of $E =\sqrt{(\alpha-\beta)C}/(4a)$ when $r=1/\sqrt{aC}$. The anisotropy $\Delta$ is a decreasing function after reaching a maximum and will have small values close to the stellar boundary. Substitution of $(\ref{M9})$ into $(\ref{M8})$ gives
\begin{equation}
\label{M10} 
4(1+ax)\ddot{y}-2a\dot{y}+(a^2-\alpha)y=0,
\end{equation}
which is the master equation.

There are three categories of solutions in terms of different values of the parameter $\alpha$. The three cases correspond to 
\begin{eqnarray}
\label{M11}
a^{2}-\alpha = 0 , ~~ a^{2}-\alpha > 0 ,~~ a^{2}-\alpha < 0,
\end{eqnarray}
which generates new models.

\section{The case $a^{2}-\alpha = 0$\label{sec3}}
With $a^{2}-\alpha = 0$, equation $(\ref{M10})$ becomes
\begin{equation}
\label{M12} 
4(1+ax)\ddot{y}-2a\dot{y}=0.
\end{equation}
Equation $(\ref{M12})$ is integrated to give
\begin{equation}
\label{M13} 
y(x)=\frac{(2+2ax)^{3/2}}{3a}c_{1}+c_{2},
\end{equation}
where $c_{1}$ and $c_{2}$ are constants.

The complete solution of the Einstein-Maxwell system is then given by
\begin{subequations}
\label{M14}
\begin{eqnarray}
\label{M14a}
e^{2\lambda}&=&1+ax,\\
\label{M14b}
e^{2\nu}&=&A^{2}\left[\frac{(2+2ax)^{3/2}}{3a}c_{1}+c_{2}\right]^{2},\\
\label{M14c}
\frac{\rho}{C}&=&\frac{6a + 2a^{2}x + (-\alpha+\beta)x}{2(1+ax)^{2}},\\
\label{M14d}
\frac{p_{r}}{C}&=&\frac{a}{1+ax} + \frac{(\alpha-\beta)x}{2(1+ax)^2}\nonumber\\ & & 
+ \frac{24ac_{1}}{4c_{1}(1+ax)^{2} + 3ac_{2}\sqrt{2(1+ax)}},\\
\label{M14e}
\frac{p_{t}}{C}&=&\frac{-2a + (-\alpha + \beta)x}{2(1+ax)^2}\nonumber\\ & &
+ \frac{12\sqrt{2}ac_{1}}{3ac_{2}\sqrt{1+ax} + 2\sqrt{2}c_{1}(1+ax)^{2}},\\
\label{M14f}
\frac{E^2}{C}&=&\frac{(\alpha - \beta)x}{(1+ax)^2},\\
\label{M14g}
\frac{\sigma^2}{C}&=&\frac{C(\alpha - \beta)(3 + ax)^2}{(1+ax)^5}.
\end{eqnarray}
\end{subequations}
The line element for this solution $(\ref{M14})$ is given by
\begin{eqnarray}
\label{M15}
ds^{2}&=&-A^{2}\left[\frac{(2+2ax)^{3/2}}{3a}c_{1}+c_{2} \right]^{2}dt^{2}\nonumber\\ & &
+ \frac{1+ax}{4Cx}dx^{2} + \frac{x}{C}(d\theta^{2} + \sin^2\theta  d\phi^{2}).
\end{eqnarray}
It is interesting to note that when $\alpha=\beta$ then the electric field vanishes and we obtain an uncharged anisotropic model.

\section{The case $a^{2}-\alpha > 0$\label{sec4}}
When $a^{2}-\alpha > 0$ then $(\ref{M10})$ has a more complicated form. However we can transform it to a standard Bessel equation. We can simplify $(\ref{M10})$ with the transformation
\begin{subequations}
\label{M16}
\begin{eqnarray} 
\label{M16a}
V&=&(1+ax)^{\frac{1}{2}},\\
\label{M16b}
y&=&Y(1+ax)^{\frac{2+a}{4}}.
\end{eqnarray}
\end{subequations}
Then $(\ref{M10})$ becomes
\begin{equation}
\label{M17}
V^{2}\frac{d^{2}Y}{dV^{2}}+V\frac{dY}{dV} 
+ \left(( a^{2}-\alpha )V^{2}-\left(\frac{2+a}{2}\right)^{2} \right)Y = 0,
\end{equation}
where $V=y^{2/(2+a)}Y^{-2/(2+a)}$. Now we use the transformation
\begin{equation}
\label{M18}
w = (a^{2}-\alpha)^{1/2}V, 
\end{equation}
to obtain
\begin{eqnarray}
\label{M19}
w^{2}\frac{d^{2}Y}{dw^{2}}+w\frac{dY}{dw}+\left(w^{2} -\left(\frac{a+2}{2}\right)^{2} \right)Y&=&0,
\end{eqnarray}
which is a Bessel equation of order $\frac{a+2}{2}$. In general the solution of $(\ref{M17})$ is a series. The general solution is a sum of linearly independent Bessel functions $J_{\frac{a+2}{2}}(w)$, of the first kind, and ${\cal{Y}}_{-\frac{a+2}{2}}(w)$, of the second kind, so that 
\begin{equation}
\label{M20}
Y(w) = b_{1}J_{\frac{a+2}{2}}(w)+b_{2}{\cal{Y}}_{-\frac{a+2}{2}}(w),
\end{equation}
and $b_{1}$, $b_{2}$ are arbitrary constants. 

The form of the solution in $(\ref{M20})$ is difficult to use in the modelling process. For specific values of $a$, when $\frac{a}{2}+1$ is a half-integer, it is possible to write the general solution of $(\ref{M20})$ as a sum of products of Legendre polynomials and trigonometric functions so that elementary functions arise. The solution has a simpler representation when $a$ is an integer. If $a=-1, 1, 3,...$ then the solution $(\ref{M20})$ can be written as Bessel functions of half-integer order $J_{\frac{1}{2}}$, $J_{-\frac{1}{2}}$, $J_{\frac{3}{2}}$, $J_{-\frac{3}{2}}$, $J_{\frac{5}{2}}$, $J_{-\frac{5}{2}}$,... (see Watson \cite{hh27}). We show that this is possible for the cases $a=-1$, $a=1$, $a=3$.

\subsection{Model I: $a=-1$}

For $a=-1$, the solution $(\ref{M20})$ can be written as
\begin{equation}
\label{M21}
Y(w) = b_{1}J_{\frac{1}{2}}(w)+b_{2}J_{-\frac{1}{2}}(w),
\end{equation}
where
\begin{subequations}
\label{M22}
\begin{eqnarray}
\label{M22a}
J_{\frac{1}{2}}(w)&=&\sqrt{\frac{2}{\pi w}}\sin(w) ,\\
\label{M22b}
J_{-\frac{1}{2}}(w)&=&-\sqrt{\frac{2}{\pi w}}\cos(w). 
\end{eqnarray}
\end{subequations}
Then the general solution of $(\ref{M10})$ is given by
\begin{eqnarray}
\label{M23}
y(x) &=& (1-\alpha)^{-1/4} 
\left[ c_{1}\sin(\sqrt{(1-\alpha)(1-x)})  \right.\nonumber\\ & & \left.
+ c_{2}\cos(\sqrt{(1-\alpha)(1-x)}) \right],
\end{eqnarray}
where $c_{1}=\sqrt{\frac{2}{\pi}}b_{1}$ and $c_{2}=-\sqrt{\frac{2}{\pi}}b_{2}$ are new constants. The complete exact solution of the Einstein-Maxwell system has the form
\begin{subequations}
\label{M24}
\begin{eqnarray}
\label{M24a}
e^{2\lambda}&=&1-x,\\
\label{M24b}
e^{2\nu}&=& (1-\alpha)^{-1/2}A^{2}
\left[ c_{1}\sin(\sqrt{(1-\alpha)(1-x)}) \right.\nonumber\\ & & \left.
+ c_{2}\cos(\sqrt{(1-\alpha)(1-x)})\right]^{2},\\
\label{M24c} 
\frac{\rho}{C}&=&\frac{-6+x(2+\beta-\alpha)}{2(1-x)^2},\\
\label{M24d}
\frac{p_{r}}{C}&=&\frac{1}{1-x}+\frac{(\alpha-\beta)x}{2(1-x)^2}\nonumber\\&&
-\left[2(1-\alpha)(c_{1}-c_{2}\tan(\sqrt{(1-\alpha)(1-x)}))\right] \nonumber\\   & & \times
\left[\sqrt{(1-\alpha)(1-x)}(c_{2}(1-x) \right.\nonumber\\&&\left.
+c_{1}(1-x)\tan(\sqrt{(1-\alpha)(1-x)}))\right]^{-1},\\
\label{M24e}
\frac{p_{t}}{C}&=&\left[4c_{1}(-1+\alpha) \right.\nonumber\\&&\left.
+ c_{2}\sqrt{(1-\alpha)(1-x)}(2+x(-2+\alpha+\beta))
\right.\nonumber\\&&\left.  + ( c_{1}\sqrt{(1-\alpha)(1-x)}(2+x(-2+\alpha+\beta)) \right.\nonumber\\&&\left.
+ 4c_{2}(1-\alpha) )\tan(\sqrt{(1-\alpha)(1-x)}) \right]\nonumber\\   & & \times
\left[ 2\sqrt{(1-\alpha)(1-x)}( c_{2}(1-x)^{2} \right.\nonumber\\&&\left.
+ c_{1}(1-x)^{2}\tan(\sqrt{(1-\alpha)(1-x)}) ) \right]^{-1},\\
\label{M24f}
\frac{E^2}{C}&=&\frac{(\alpha-\beta)x}{(1-x)^{2}},\\
\label{M24g}
\frac{\sigma^2}{C}&=&\frac{C(\alpha-\beta)(-3+x)^{2}}{(1-x)^{5}}.
\end{eqnarray}
\end{subequations}

This is a new solution to the Einstein-Maxwell system. The line element for this case is
\begin{eqnarray}
\label{M25}
ds^2&=&-(\alpha-1)^{-1/2}A^{2} 
\left[ c_{1}\sin(\sqrt{(1-\alpha)(1-x)}) \right.\nonumber\\&&\left.
+ c_{2}\cos(\sqrt{(1-\alpha)(1-x)})\right]^{2}dt^{2} \nonumber\\ &&
+\frac{1-x}{4Cx}dx^{2} + \frac{x}{C}(d\theta^{2} + \sin^{2}\theta d\phi^{2}).
\end{eqnarray}

\subsection{Model II: $a=1$}
\label{Model II: $a=1$}
When $a=1$, $(\ref{M20})$ is of the form
\begin{equation}
\label{M26}
Y(w) = b_{1}J_{\frac{3}{2}}(w)+b_{2}J_{-\frac{3}{2}}(w),
\end{equation}
where 
\begin{subequations}
\label{M27}
\begin{eqnarray}
\label{M27a}
J_{\frac{3}{2}}(w)&=&\sqrt{\frac{2}{\pi w}}\left[\frac{\sin(w)}{w} - \cos(w)\right],\\
\label{M27b}
J_{-\frac{3}{2}}(w)&=&-\sqrt{\frac{2}{\pi w}}\left[\frac{\cos(w)}{w} + \sin(w)\right].
\end{eqnarray}
\end{subequations}
Then the general solution to $(\ref{M10})$ is
\begin{eqnarray}
\label{M28}
y(x)&=&(1-\alpha)^{-3/4} 
\left[c_{2}\cos\left(\sqrt{(1-\alpha)(1+x)}\right) \right.\nonumber\\&& \left.
 - c_{1}\sqrt{(1-\alpha)(1+x)}\cos\left(\sqrt{(1-\alpha)(1+x)}\right) \right.\nonumber\\&& \left. 
+ c_{2}\sqrt{(1-\alpha)(1+x)}\sin\left(\sqrt{(1-\alpha)(1+x)}\right) \right.\nonumber\\&& \left.
+ c_{1}\sin\left(\sqrt{(1-\alpha)(1+x)}\right)\right],
\end{eqnarray}
where we introduced the constants $c_{1}=\sqrt{\frac{2}{\pi}}b_{1}$ and $c_{2}=-\sqrt{\frac{2}{\pi}}b_{2}$. This form of solution is similar to previous studies. With the help of the general solution ($\ref{M28}$), we can write the complete exact charged anisotropic solution of the Einstein-Maxwell system as 
\begin{subequations}
\label{M29}
\begin{eqnarray}
\label{M29a}
e^{2\lambda}&=&1+x,\\
\label{M29b}
e^{2\nu}&=& (1-\alpha)^{-3/2}A^{2} 
\left[c_{2}\cos\left(\sqrt{(1-\alpha)(1+x)}\right) \right.\nonumber\\&& \left.
 - c_{1}\sqrt{(1-\alpha)(1+x)}\cos\left(\sqrt{(1-\alpha)(1+x)}\right) \right.\nonumber\\&& \left. 
+ c_{2}\sqrt{(1-\alpha)(1+x)}\sin\left(\sqrt{(1-\alpha)(1+x)}\right) \right.\nonumber\\&& \left.
+ c_{1}\sin\left(\sqrt{(1-\alpha)(1+x)}\right)\right]^{2},\\
\label{M29c} 
\frac{\rho}{C}&=&\frac{6+x(2+\beta-\alpha)}{2(1+x)^2},\\
\label{M29d}
\frac{p_{r}}{C}&=&\left[\frac{c_{1}(2+(2-\alpha+\beta)x)\sqrt{(1-\alpha)(1+x)}}{\tan(\sqrt{(1-\alpha)(1+x)})} \right.\nonumber\\&& \left.
- \frac{c_{2}(-2+4\alpha + (-2+3\alpha + \beta)x)}{\tan(\sqrt{(1-\alpha)(1+x)})} \right.\nonumber\\&& \left.
- c_{2}(2+(2-\alpha+\beta)x)\sqrt{(1-\alpha)(1+x)} \right.\nonumber\\&& \left. 
- c_{1}(-2+4\alpha + (-2+3\alpha + \beta)x)\right] \nonumber\\   && \times
\left[ \frac{2(1+x)^{2}(c_{2}-c_{1} (\sqrt{(1-\alpha)(1+x)}))}{\tan(\sqrt{(1-\alpha)(1+x)})} \right.\nonumber\\&& \left.
+ 2(1+x)^{2}
(c_{1} + c_{2}\sqrt{(1-\alpha)(1+x)})\right]^{-1}, \\ 
\label{M29e}
\frac{p_{t}}{C}&=&\left[\frac{c_{1}(2+(2+\alpha - \beta)x)\sqrt{(1-\alpha)(1+x)}}{\tan(\sqrt{(1-\alpha)(1+x)})} \right.\nonumber\\&& \left.
+ \frac{c_{2}(2 -4\alpha + (2-3\alpha + \beta)x)}{\tan(\sqrt{(1-\alpha)(1+x)})} \right.\nonumber\\&& \left.
+ c_{1}(2-4\alpha + (2-3\alpha + \beta)x) \right.\nonumber\\&& \left.
+ c_{2}(-2+(-2-\alpha+\beta)x)\sqrt{(1-\alpha)(1+x)} \right] \nonumber\\   && \times
\left[\frac{2(1+x)^{2}(c_{2}-c_{1} (\sqrt{(1-\alpha)(1+x)}))}{\tan(\sqrt{(1-\alpha)(1+x)})} \right.\nonumber\\&& \left.
+ 2(1+x)^{2} (c_{1} + c_{2}\sqrt{(1-\alpha)(1+x)})\right]^{-1},\\ 
\label{M29f} 
\frac{\ E^2}{C}&=&\frac{(\alpha-\beta)x}{(1+x)^2},\\
\label{M29g}
\frac{\sigma^2}{C}&=&\frac{C(3+x)^{2}(\alpha-\beta)}{(1+x)^{5}}.
\end{eqnarray}
\end{subequations}
The system $(\ref{M29})$ gives the exact solution of the Einstein-Maxwell system expressed in terms of elementary functions. This is a new solution. We can consider the result  $(\ref{M29})$ as a generalisation of the Hansraj and Maharaj \cite{hh14} model; when $\beta=0$ the pressures are isotropic and we regain their model. When $\alpha=0$ and $\beta=0$ then we have an uncharged isotropic star which was the model first found by Finch and Skea \cite{hh13}. We can write the line element in terms of the coordinate $x$ as
\begin{eqnarray}
\label{M30}
ds^{2}&=& -(1-\alpha)^{-3/2}A^{2}
\left[c_{2}\cos(\sqrt{(1-\alpha)(1+x)}) \right.\nonumber\\&& \left.   
- c_{1}\sqrt{(1-\alpha)(1+x)}\cos(\sqrt{(1-\alpha)(1+x)}) \right.\nonumber\\&& \left. 
+ c_{2}\sqrt{(1-\alpha)(1+x)}\sin(\sqrt{(1-\alpha)(1+x)}) \right.\nonumber\\&& \left. 
+ c_{1}\sin(\sqrt{(1-\alpha)(1+x)})\right]^{2}dt^{2}\nonumber\\& &
+ \frac{1+x}{4Cx}dx^{2}+\frac{x}{C}(d\theta^{2}+\sin^{2}\theta d\phi^{2}).
\end{eqnarray}
The metric $(\ref{M30})$ may be interpreted as the anisotropic, charged generalisation of the Finch and Skea \cite{hh13} solution.

\subsection{Model III: $a=3$}

When $a=3$, $(\ref{M20})$ is of the form
\begin{equation}
\label{M31}
Y(w) = b_{1}J_{\frac{5}{2}}(w)+b_{2}J_{-\frac{5}{2}}(w),
\end{equation}
where
\begin{subequations}
\label{M32}
\begin{eqnarray}
\label{M32a}
J_{\frac{ 5}{2}}(w)&=&\sqrt{\frac{2}{\pi w}}\left(\frac{3\sin w}{w^2}-\frac{3\cos w}{w}-\sin w\right),\\
\label{M32b}
J_{-\frac{ 5}{2}}(w)&=&\sqrt{\frac{2}{\pi w}}
\left(-\frac{3\cos w}{w^2}-\frac{3\sin w}{w}+\cos w\right).
\end{eqnarray}
\end{subequations}
Then the general solution to $(\ref{M10})$ is
\begin{eqnarray}
\label{M33}
y(x)&=&(9-\alpha)^{-5/4}
\left[-3c_{1}\sqrt{(9-\alpha)(1+3x)} \cos\sqrt{(9-\alpha)(1+3x)} \right.\nonumber\\&& \left.
+ c_{2}(9-\alpha)(1+3x)\cos\sqrt{(9-\alpha)(1+3x)}\right.\nonumber\\& &\left. 
- 3c_{2}\cos\sqrt{(9-\alpha)(1+3x)} \right.\nonumber\\&& \left.
- 3c_{2}\sqrt{(9-\alpha)(1+3x)} \sin\sqrt{(9-\alpha)(1+3x)} \right.\nonumber\\&& \left.
- c_{1}(9-\alpha)(1+3x)\sin\sqrt{(9-\alpha)(1+3x)} \right.\nonumber\\&& \left.
+ 3c_{1} \sin\sqrt{(9-\alpha)(1+3x)}\right],
\end{eqnarray}
where we have defined $c_{1}=a\sqrt{\frac{2}{\pi}}$ and $c_{2}=b\sqrt{\frac{2}{\pi}}$ as new constants. The complete exact solution to the Einstein-Maxwell system for this case is thus given by
\begin{subequations}
\label{M34}
\begin{eqnarray}
\label{M34a}
e^{2\lambda}&=&1+3x,\\
\label{M34b}
e^{2\nu}&=&\frac{ A^2}{(9-\alpha)^{5/2}}
\left[-3c_{1}\sqrt{(9-\alpha)(1+3x)} \cos\sqrt{(9-\alpha)(1+3x)} \right.\nonumber\\&& \left.
+ c_{2}(9-\alpha)(1+3x)\cos\sqrt{(9-\alpha)(1+3x)}\right.\nonumber\\& &\left. 
- 3c_{2}\cos\sqrt{(9-\alpha)(1+3x)} \right.\nonumber\\&& \left.
- 3c_{2}\sqrt{(9-\alpha)(1+3x)} \sin\sqrt{(9-\alpha)(1+3x)} \right.\nonumber\\&& \left.
- c_{1}(9-\alpha)(1+3x)\sin\sqrt{(9-\alpha)(1+3x)} \right.\nonumber\\&& \left.
+ 3c_{1} \sin\sqrt{(9-\alpha)(1+3x)}\right]^{2},\\
\label{M34c}
\frac{\rho}{C}&=&\frac{18+\left(18-\alpha+\beta\right)x}{2(1+3x)^2},\\
\label{M34d}
\frac{ p_{r}}{C}&=& 6 \left(9-\alpha\right)(1+3x)^{-1}
\left[-c_{2}-c_{1}\sqrt{(9-\alpha)(1+3x)} + (c_{1} \right.\nonumber\\&& \left.
-c_{2}\sqrt{(9-\alpha)(1+3x)})\tan(\sqrt{(9-\alpha)(1+3x)}) \right] \nonumber\\&& \times
\left[-3c_{1}\sqrt{(9-\alpha)(1+3x)} \right.\nonumber\\&& \left.
 +  c_{2}((9-\alpha)(1+3x)-3) \right.\nonumber\\&& \left.
 -3c_{2}\sqrt{(9-\alpha)(1+3x)}\tan(\sqrt{(9-\alpha)(1+3x)}) \right.\nonumber\\&& \left.
-c_{1}(9-\alpha)(1+3x)\tan(\sqrt{(9-\alpha)(1+3x)})\right.\nonumber\\&& \left.
+ 3c_{1}\tan(\sqrt{(9-\alpha)(1+3x)})\right]^{-1} \nonumber\\&&
- \frac{6+(18-\alpha+\beta)x}{12(9-\alpha)(1+3x)}, \\
\label{M34e}
\frac{p_{t}}{C}&=& 6 \left(9-\alpha\right)(1+3x)^{-1}
\left[-c_{2}-c_{1}\sqrt{(9-\alpha)(1+3x)} + (c_{1} \right.\nonumber\\&& \left.
- c_{2}\sqrt{(9-\alpha)(1+3x)})\tan(\sqrt{(9-\alpha)(1+3x)}) \right] \nonumber\\&& \times
\left[-3c_{1}\sqrt{(9-\alpha)(1+3x)} \right.\nonumber\\&& \left.
 +  c_{2}((9-\alpha)(1+3x)-3) \right.\nonumber\\&& \left.
 -3c_{2}\sqrt{(9-\alpha)(1+3x)}\tan(\sqrt{(9-\alpha)(1+3x)}) \right.\nonumber\\&& \left.
-c_{1}(9-\alpha)(1+3x) \tan(\sqrt{(9-\alpha)(1+3x)}) \right.\nonumber\\&& \left.
+ 3c_{1}\tan(\sqrt{(9-\alpha)(1+3x)})\right]^{-1} \nonumber\\&&
- \frac{6+(18-\alpha-\beta)x}{12(9-\alpha)(1+3x)}, \\
\label{M34f} 
\frac{E^2}{C}&=&\frac{(\alpha-\beta)x}{(1+3x)^2},\\
\label{M34g}
\frac{\sigma^2}{C}&=&\frac{9C(\alpha-\beta)(1+x)^{2}}{(1+3x)^5}.
\end{eqnarray}
\end{subequations}

This is a new category of exact models for a charged, anisotropic matter distribution. The line element is given by
\begin{eqnarray}
\label{M35}
ds^{2}&=&-A^{2}(9-\alpha)^{-5/2}
\left[-3c_{1}\sqrt{(9-\alpha)(1+3x)} \cos\sqrt{(9-\alpha)(1+3x)} \right.\nonumber\\&& \left.
+ c_{2}(9-\alpha)(1+3x)\cos\sqrt{(9-\alpha)(1+3x)}\right.\nonumber\\& &\left. 
- 3c_{2}\cos\sqrt{(9-\alpha)(1+3x)} \right.\nonumber\\&& \left.
- 3c_{2}\sqrt{(9-\alpha)(1+3x)} \sin\sqrt{(9-\alpha)(1+3x)} \right.\nonumber\\&& \left.
- c_{1}(9-\alpha)(1+3x)\sin\sqrt{(9-\alpha)(1+3x)} \right.\nonumber\\&& \left.
+ 3c_{1} \sin\sqrt{(9-\alpha)(1+3x)}\right]^{2}dt^{2}\nonumber\\ &&
+\frac{1+3x}{4Cx}dx^{2}+\frac{x}{C}(d\theta^{2}+\sin^{2}\theta d\phi^{2}).
\end{eqnarray}

\section{The case $a^{2}-\alpha < 0$\label{sec5}}
We now consider the case $a^{2}-\alpha < 0$ and write the differential equation $(\ref{M10})$ as
\begin{equation}
\label{M36} 
4(1+ax)\ddot{y}-2a\dot{y}-(\alpha-a^2)y=0.
\end{equation}
Keeping the same transformation $(\ref{M16})$ of Sect.~\ref{sec4}, the equation $(\ref{M36})$ takes the form
\begin{equation}
\label{M37}
V^{2}\frac{d^{2}Y}{dV^{2}}+V\frac{dY}{dV} 
- \left((\alpha - a^{2})V^{2} + \left(\frac{2+a}{2}\right)^{2} \right)Y = 0,
\end{equation}
where $V=y^{2/(2+a)}Y^{-2/(2+a)}$. We cannot use the variable $w$ of Sect.~\ref{sec4} as $a^{2}-\alpha < 0$. It is important to use a new variable $\tilde{w}$. By taking
\begin{equation}
\label{M38}
\tilde{w}=(\alpha - a^{2})^{\frac{1}{2}}V,
\end{equation}
equation $(\ref{M36})$ becomes
\begin{equation}
\label{M39}
\tilde{w}^{2}\frac{d^{2}Y}{d\tilde{w}^{2}} + \tilde{w}\frac{dY}{d\tilde{w}} - \left( \tilde{w}^{2} + \left(\frac{2+a}{2}\right)^{2} \right)Y = 0.
\end{equation}
Equation $(\ref{M39})$ is the modified Bessel differential equation of order $\frac{2+a}{2}$. The general solution of $(\ref{M39})$ is a sum of linearly independent modified Bessel functions given by
\begin{equation}
\label{M40}
Y(\tilde{w}) = b_{1}I_{\frac{a+2}{2}}(\tilde{w})+b_{2}K_{-\frac{a+2}{2}}(\tilde{w}),
\end{equation}
where $b_{1}$, $b_{2}$ are arbitrary constants. The quantities $I_{\frac{a+2}{2}}(\tilde{w})$, $K_{-\frac{a+2}{2}}(\tilde{w})$ are called modified Bessel functions of the first and second kind respectively. The form of the solution of $(\ref{M40})$ is complicated but can be written in terms of elementary functions when $\frac{a}{2}+1$ is a half-integer. For these parameter values the solution is usually written in terms of hyperbolic functions. For $a=-1, 1, 3,...$ the solution of $(\ref{M39})$ can be written with the help of modified Bessel functions of half-integer order $I_{\frac{1}{2}}$, $I_{-\frac{1}{2}}$, $I_{\frac{3}{2}}$, $I_{-\frac{3}{2}}$, $I_{\frac{5}{2}}$, $I_{-\frac{5}{2}}$,... We now consider the cases where $a=-1$, $a=1$ and $a=3$.

\subsection{Model I: $a=-1$}

When $a=-1$ the solution $(\ref{M40})$ takes the form
\begin{equation}
\label{M41}
Y(\tilde{w}) = b_{1}I_{\frac{1}{2}}(\tilde{w})+b_{2}I_{-\frac{1}{2}}(\tilde{w}),
\end{equation}
where
\begin{subequations}
\label{M42}
\begin{eqnarray}
\label{M42a}
I_{\frac{1}{2}}(\tilde{w})&=&\sqrt{\frac{2}{\pi \tilde{w}}}\sinh(\tilde{w}) ,\\
\label{M42b}
I_{-\frac{1}{2}}(\tilde{w})&=&\sqrt{\frac{2}{\pi \tilde{w}}}\cosh(\tilde{w}). 
\end{eqnarray}
\end{subequations}
Then the general solution of $(\ref{M36})$ is given by
\begin{eqnarray}
\label{M43}
y(x)&=& (\alpha - 1)^{-1/4} 
\left[ c_{1}\sinh(\sqrt{(\alpha-1)(1-x)}) \right.\nonumber\\ & & \left.
 + c_{2}\cosh(\sqrt{(\alpha-1)(1-x)}) \right],
\end{eqnarray}
where $c_{1}=\sqrt{\frac{2}{\pi}}b_{1}$ and $c_{2}=\sqrt{\frac{2}{\pi}}b_{2}$ are new constants. Then the complete exact solution of the Einstein-Maxwell system is
\begin{subequations}
\label{M44}
\begin{eqnarray}
\label{M44a}
e^{2\lambda}&=&1-x,\\
\label{M44b}
e^{2\nu}&=& (\alpha-1)^{-1/2}A^{2}
\left[ c_{1}\sinh(\sqrt{(\alpha-1)(1-x)}) \right.\nonumber\\ & & \left.
+ c_{2}\cosh(\sqrt{(\alpha-1)(1-x)})\right]^{2},\\
\label{M44c} 
\frac{\rho}{C}&=&\frac{-6+x(2+\beta-\alpha)}{2(1-x)^2},\\
\label{M44d}
\frac{p_{r}}{C}&=&\frac{1}{1-x}+\frac{(\alpha-\beta)x}{2(1-x)^2}\nonumber\\&&
+\left[2(1-\alpha)(c_{1}+c_{2}\tanh(\sqrt{(\alpha-1)(1-x)}))\right] \nonumber\\   & & \times
\left[\sqrt{(\alpha-1)(1-x)}(c_{2}(1-x) \right.\nonumber\\&&\left.
+c_{1}(1-x)\tanh(\sqrt{(\alpha-1)(1-x)}))\right]^{-1},\\
\label{M44e}
\frac{p_{t}}{C}&=&\left[-4c_{1}(-1+\alpha) \right.\nonumber\\&&\left.
 + c_{2}\sqrt{(\alpha-1)(1-x)}(2+x(-2+\alpha+\beta))
\right.\nonumber\\&&\left.  + ( c_{1}\sqrt{(\alpha-1)(1-x)}(2+x(-2+\alpha+\beta)) \right.\nonumber\\&&\left.
- 4c_{2}(-1+\alpha) )\tanh(\sqrt{(\alpha-1)(1-x)}) \right]\nonumber\\   & & \times
\left[ 2\sqrt{(\alpha-1)(1-x)}( c_{2}(1-x)^{2} \right.\nonumber\\&&\left.
+ c_{1}(1-x)^{2}\tanh(\sqrt{(\alpha-1)(1-x)}) ) \right]^{-1},\\
\label{M44f}
\frac{E^2}{C}&=&\frac{(\alpha-\beta)x}{(1-x)^{2}},\\
\label{M44g}
\frac{\sigma^2}{C}&=&\frac{C(\alpha-\beta)(-3+x)^{2}}{(1-x)^{5}}.
\end{eqnarray}
\end{subequations}

This is a new solution to the Einstein-Maxwell system in terms of hyperbolic functions. The line element for this case is
\begin{eqnarray}
\label{M45}
ds^2&=&-(\alpha-1)^{-1/2}A^{2}
\left[ c_{1}\sinh(\sqrt{(\alpha-1)(1-x)}) \right.\nonumber\\&&\left.
+c_{2}\cosh(\sqrt{(\alpha-1)(1-x)})\right]^{2}dt^{2} \nonumber\\ &&
+\frac{1-x}{4Cx}dx^{2} + \frac{x}{C}(d\theta^{2} + \sin^{2}\theta d\phi^{2}).
\end{eqnarray}

\subsection{Model II: $a=1$}

For $a=1$ the solution $(\ref{M40})$ becomes
\begin{equation}
\label{M46}
Y(\tilde{w}) = b_{1}I_{\frac{3}{2}}(\tilde{w})+b_{2}I_{-\frac{3}{2}}(\tilde{w}),
\end{equation}
where the modified Bessel functions are given by
\begin{subequations}
\label{M47}
\begin{eqnarray}
\label{M47a}
I_{\frac{3}{2}}(\tilde{w})&=&\sqrt{\frac{2}{\pi\tilde{w}}}\left[-\frac{\sinh(\tilde{w})}{\tilde{w}} + \cosh(\tilde{w})\right],\\
\label{M47b}
I_{-\frac{3}{2}}(\tilde{w})&=&\sqrt{\frac{2}{\pi\tilde{w}}}\left[-\frac{\cosh(\tilde{w})}{\tilde{w}} + \sinh(\tilde{w})\right],
\end{eqnarray}
\end{subequations}
Then the general solution of the equation $(\ref{M36})$ takes the form
\begin{eqnarray}
\label{M48}
y(x)&=&(\alpha-1)^{-\frac{3}{4}} 
\left[c_{1}\sqrt{(\alpha-1)(1+x)}\sinh(\sqrt{(\alpha-1)(1+x)}) \right.\nonumber\\&& \left.
- c_{2}\sinh(\sqrt{(\alpha-1)(1+x)}) \right.\nonumber\\&& \left.
+ c_{2}\sqrt{(\alpha-1)(1+x)}\cosh(\sqrt{(\alpha-1)(1+x)}) \right.\nonumber\\&& \left.
- c_{1}\cosh(\sqrt{(\alpha-1)(1+x)})\right],
\end{eqnarray}
where $c_{1}=b_{2}\sqrt{\frac{2}{\pi}}$ and $c_{2}=b_{1}\sqrt{\frac{2}{\pi}}$ are new constants. The complete exact solution to the Einstein-Maxwell system for this case can be written as
\begin{subequations}
\label{M49}
\begin{eqnarray}
\label{M49a}
e^{2\lambda}&=&1+x,\\
\label{M49b}
e^{2\nu}&=& (\alpha - 1)^{-3/2}A^{2}
\left[c_{1}\sqrt{(\alpha-1)(1+x)}\sinh(\sqrt{(\alpha-1)(1+x)}) \right.\nonumber\\&& \left.
- c_{2}\sinh(\sqrt{(\alpha-1)(1+x)}) \right.\nonumber\\&& \left.
+ c_{2}\sqrt{(\alpha-1)(1+x)}\cosh(\sqrt{(\alpha-1)(1+x)}) \right.\nonumber\\&& \left.
- c_{1}\cosh(\sqrt{(\alpha-1)(1+x)})\right]^{2},\\
\label{M49c}
\frac{\rho}{C}&=&\frac{6 - (\alpha-1)x + (1+\beta)x}{2(1+x)^{2}},\\
\label{M49d}
\frac{p_{r}}{C}&=&\left[c_{2}\sqrt{(\alpha-1)(1+x)}(2+x(2-\alpha+\beta)) \right.\nonumber\\& &\left.  
- c_{1}(-2+4\alpha+x(-2+3\alpha+\beta))  \right.\nonumber\\& &\left. 
+ (c_{1}\sqrt{(\alpha-1)(1+x)}(2+x(2-\alpha+\beta))  \right.\nonumber\\& &\left.
- c_{2}(-2+4\alpha+x(-2+3\alpha \right.\nonumber\\& &\left.
+\beta)))\tanh(\sqrt{(\alpha-1)(1+x)})\right]\nonumber\\ && \times
\left[2(1+x)^{2}(c_{1}-c_{2}\sqrt{(\alpha-1)(1+x)} \right.\nonumber\\& &\left.
+ c_{2}\tanh(\sqrt{(\alpha-1)(1+x)}) \right.\nonumber\\& &\left.
-c_{1}\sqrt{(\alpha-1)(1+x)} \right.\nonumber\\   
 &&\left. \times \tanh(\sqrt{(\alpha-1)(1+x)}))\right]^{-1},\\
\label{M49e}
\frac{p_{t}}{C}&=&\left[c_{1}(2-4\alpha+x(2-3\alpha+\beta)) \right.\nonumber\\& &\left. 
- c_{2}\sqrt{(\alpha-1)(1+x)}(-2+x(-2+\alpha+\beta)) \right.\nonumber\\& &\left.
+ (c_{2}(2-4\alpha+x(2-3\alpha+\beta)) \right.\nonumber\\& &\left. 
- c_{1}\sqrt{(\alpha-1)(1+x)}(x(-2+\alpha+\beta) \right.\nonumber\\& &\left. 
-2))\tanh(\sqrt{(\alpha-1)(1+x)})\right] \nonumber\\ && \times
\left[2(1+x)^{2}(c_{1}-c_{2}\sqrt{(\alpha-1)(1+x)} \right.\nonumber\\& &\left.
+c_{2}\tanh(\sqrt{(\alpha-1)(1+x)}) \right.\nonumber\\& &\left.
- c_{1}\sqrt{(\alpha-1)(1+x)} \right.\nonumber\\   
 &&\left. \times \tanh(\sqrt{(\alpha-1)(1+x)}))\right]^{-1},\\
\label{M49f} 
\frac{E^2}{C}&=&\frac{(\alpha-1)x + (1-\beta)x}{(1+ax)^2},\\
\label{M49g}
\frac{\sigma^2}{C}&=&\frac{C((1-\beta) + (\alpha-1) )(3 + ax)^2}{(1+ax)^5}.
\end{eqnarray}
\end{subequations}

Equations $(\ref{M49})$ represent a new solution in terms of hyperbolic functions. This result is a generalisation of the corresponding metric of Hansraj and Maharaj \cite{hh14}; when $\beta=0$ the anisotropy vanishes and we regain their model. The line element takes the form
\begin{eqnarray}
\label{M50}
ds^{2}&=&-A^{2}(\alpha-1)^{-\frac{3}{2}}
\left[c_{1}\sqrt{(\alpha-1)(1+x)}\sinh(\sqrt{(\alpha-1)(1+x)}) \right.\nonumber\\&& \left.
- c_{2}\sinh(\sqrt{(\alpha-1)(1+x)}) \right.\nonumber\\&& \left.
+ c_{2}\sqrt{(\alpha-1)(1+x)}\cosh(\sqrt{(\alpha-1)(1+x)}) \right.\nonumber\\&& \left.
- c_{1}\cosh(\sqrt{(\alpha-1)(1+x)})\right]^{2}dt^{2}\nonumber\\ & &
+ \frac{1+x}{4Cx}dx^{2} + \frac{x}{C}(d\theta^{2} + \sin^{2}\theta d\phi^{2}).
\end{eqnarray}

\subsection{Model III: $a=3$}

When $a=3$ we can write the solution $(\ref{M40})$ as
\begin{equation}
\label{M51}
Y(\tilde{w}) = b_{1}I_{\frac{5}{2}}(\tilde{w})+b_{2}I_{-\frac{5}{2}}(\tilde{w}),
\end{equation}
where $b_{1}$ , $b_{2}$ are constants and $I_{\frac{ 5}{2}}$ , $I_{-\frac{ 5}{2}}$ are modified Bessel functions which may be expressed in terms of hyperbolic functions as 
\begin{subequations}
\label{M52}
\begin{eqnarray}
\label{M52a}
I_{\frac{5}{2}}(\tilde{w})&=&\sqrt{\frac{2}{\pi \tilde{w}}} 
\left(\frac{3\sinh(\tilde{w})}{\tilde{w}^2} 
-\frac{3\cosh(\tilde{w})}{\tilde{w}}+\sinh(\tilde{w})\right),\\
\label{M52b}
I_{-\frac{5}{2}}(\tilde{w})&=&\sqrt{\frac{2}{\pi \tilde{w}}} 
\left(\frac{3\cosh(\tilde{w})}{\tilde{w}^2} 
-\frac{3\sinh(\tilde{w})}{w}+\cosh(\tilde{w})\right).
\end{eqnarray}
\end{subequations}
Then the general solution to the differential equation in this case may be written as
\begin{eqnarray}
\label{M53}
y(x)&=&(\alpha-9)^{-5/4}
\left[((3+(\alpha-9)(1+3x))c_{1}\right.\nonumber\\& &\left.
-3c_{2}\sqrt{(\alpha-9)(1+3x)})\sinh(T(x)) \right.\nonumber\\& &\left.
 + ((3+(\alpha-9)(1+3x))c_{2}\right.\nonumber\\& &\left. 
 -3c_{1}\sqrt{(\alpha-9)(1+3x)})\cosh(T(x)) \right],
\end{eqnarray}
where $c_{1}=b_{1}\sqrt{\frac{2}{\pi}} $ and $c_{2}=b_{2}\sqrt{\frac{2}{\pi}} $ are new constants. The complete solution to the Einstein-Maxwell equations is given by
\begin{subequations}
\label{M54}
\begin{eqnarray}                                   
\label{M54a}
e^{2\lambda}&=&1+3x,\\
\label{M54b}
e^{2\nu}&=& A^{2}(\alpha-9)^{-5/2}
\left[((3+(\alpha-9)(1+3x))c_{1}\right.\nonumber\\& &\left.
-3c_{2}\sqrt{(\alpha-9)(1+3x)})\sinh(T(x)) \right.\nonumber\\& &\left.
 + ((3+(\alpha-9)(1+3x))c_{2}\right.\nonumber\\& &\left. 
 -3c_{1}\sqrt{(\alpha-9)(1+3x)})\cosh(T(x))\right]^{2},\\
\label{M54c} 
\frac{\rho}{C}&=&\frac{18+\left(18-\alpha+\beta\right)x}{2(1+3x)^2},\\
\label{M54d}
\frac{p_{r}}{C}&=&\frac{-6+(-18+\alpha-\beta)x}{2(1+3x)^2} 
\left[ 12(1+3x)(\alpha-9)(-c_{2}+c_{1}T(x)    \right.\nonumber\\& &\left.
+(-c_{1}+c_{1}T(x))\tanh(T(x))) \right]\nonumber\\   & & \times
\left[2(1+3x)^{2}(-3c_{1}T(x) \right.\nonumber\\& &\left.
+ c_{2}(3+(\alpha-9)(1+3x)) + (-3c_{2}T(x) \right.\nonumber\\& &\left.
+ c_{1}(3+(\alpha-9)(1+3x))) \right.\nonumber\\   
 &&\left. \times  \tanh(T(x)))\right]^{-1},\\
\label{M54e}
\frac{p_{t}}{C}&=& \left[ -(3c_{1}(1+3x)(\alpha-9)(-30+4\alpha \right.\nonumber\\& &\left.
+x(-54+7\alpha-\beta)) \right.\nonumber\\& &\left.
+c_{2}T(x)(-18(-8+\alpha)+x(-6(-297+\beta) \right.\nonumber\\& &\left.
+\alpha(-354+17\alpha+\beta) \right.\nonumber\\& &\left.
+3x(\alpha-9)(-162+17\alpha+\beta)))) \right.\nonumber\\& &\left.
-(3c_{2}(\alpha-9)(1+3x)(-30+4\alpha \right.\nonumber\\& &\left.
+x(-54+7\alpha-\beta))  \right.\nonumber\\& &\left.
+c_{1}T(x)(-18(-8+\alpha)+x(-6(-297+\beta) \right.\nonumber\\& &\left.
+\alpha(-354+17\alpha+\beta)  \right.\nonumber\\& &\left.
+3x(\alpha-9)(-162+17\alpha+\beta))))\tanh(T(x)) \right]   \nonumber\\   & & \times 
\left[2(1+3x)^{2}T(x) (3c_{1}T(x) \right.\nonumber\\& &\left.
	- c_{2}(-6+3x(\alpha-9)+\alpha)	+ (	3c_{2}T(x)	\right.\nonumber\\& &\left.
	- c_{1}(-6+3x(\alpha-9)+\alpha)) \right.\nonumber\\   
 &&\left. \times \tanh(T(x))) \right]^{-1},   \\
\label{M54f} 
\frac{E^2}{C}&=&\frac{\left(9-(9-\alpha)-\beta\right)x}{(1+3x)^2},\\
\label{M54g}
\frac{\sigma^2}{C}&=&\frac{\ 9C\left(9-(9-\alpha)-\beta\right)(1+x)^{2}}{(1+3x)^5},
\end{eqnarray}
\end{subequations}
where we have set $T(x) \equiv \sqrt{(\alpha-9)(1+3x)}.$

We have found another class of new solutions which allows for more complex behaviour in the potentials than the earlier cases. For our new solution to the Einstein-Maxwell system given by equations $(\ref{M54})$, the line element has the form 
\begin{eqnarray}
\label{M55}
ds^{2}&=&-A^{2}(\alpha-9)^{-5/2}
\left[((3+(\alpha-9)(1+3x))c_{1}\right.\nonumber\\& &\left.
-3c_{2}T(x))\sinh(T(x)) \right.\nonumber\\& &\left.
 + ((3+(\alpha-9)(1+3x))c_{2}\right.\nonumber\\& &\left. 
 -3c_{1}T(x))\cosh(T(x))\right]^{2}dt^{2} \nonumber\\ &&
+ \frac{1+3x}{4Cx}dx^{2}+\frac{x}{C}(d\theta^{2} + \sin^{2}\theta d\phi^{2}).
\end{eqnarray}

\section{Equation of state\label{sec6}}

An equation of state relating the radial pressure $p_{r}$ to the energy density $\rho$ is a desirable physical feature in a relativistic stellar model. The expressions for the radial pressure $p_{r}$ are complicated but all the models found in this paper admit an equation of state. We illustrate this for the model found in Sect.~\ref{Model II: $a=1$}. From equation $(\ref{M29c})$ in Sect.~\ref{Model II: $a=1$} we can establish the expression
\begin{equation}
\label{M56}
x^{2}+\frac{4\rho - C(2+\beta-\alpha)}{2\rho}x+\frac{\rho-3C}{\rho} = 0.
\end{equation}
To solve this equation, with distinct real roots, we impose the following condition, where the discriminant of the quadratic equation
$(\ref{M56})$ is positive. We have that
\begin{eqnarray}
\label{M57}
\left(\frac{C}{2\rho}(2+\beta-\alpha)\right)^{2}+\frac{2C}{\rho}(4-\beta+\alpha)&>&0.
\end{eqnarray}
Hence the variable $x$ is written in terms of $\rho$ as
\begin{eqnarray}
\label{M58} 
x &=& -\left[1-\frac{C}{4\rho}(2+\beta-\alpha) \right] \nonumber\\&&
+\frac{1}{2}\sqrt{\left(\frac{C}{2\rho}(2+\beta-\alpha) \right)^{2}+\frac{2C}{\rho}(4-\beta+\alpha)}.
\end{eqnarray}
Then from $(\ref{M29d})$ we can write $p_{r}$ as a function of $\rho$. Therefore, we have an equation of state of the form
\begin{eqnarray}
\label{M59}
\frac{p_{r}}{C}&=&\left[-c_{2}\left(-2+4\alpha+(-2+3\alpha+\beta)(-1+F(\rho))\right)\right.\nonumber\\& &\left.
+ c_{1}(2+(2-\alpha+\beta)(-1+F(\rho)) )\sqrt{(1-\alpha)F(\rho)} \right.\nonumber\\& &\left.
-( c_{1}(-2+4\alpha+(-2+3\alpha+\beta)(-1+F(\rho))) \right.\nonumber\\& &\left.
 + c_{2}(2+(2-\alpha +\beta)(-1        \right.\nonumber\\& &\left.
 + F(\rho)) )\sqrt{(1-\alpha)F(\rho)} )\tan(\sqrt{(1-\alpha)F(\rho)}) \right] \nonumber\\& & \times
\left[2(F(\rho))^{2} ( c_{2}-c_{1}\sqrt{(1-\alpha)F(\rho)} +(c_{1}     \right.\nonumber\\& &\left.
 + c_{2}\sqrt{(1-\alpha)F(\rho)} )\tan(\sqrt{(1-\alpha)F(\rho)}) )\right]^{-1}, 
\end{eqnarray}
where we have set
\begin{eqnarray}
\label{M60}
F(\rho)&=&\frac{1}{2}\sqrt{\left(\frac{C}{2\rho}(2-\alpha+\beta)\right)^{2}+\frac{2C}{\rho}(4+\alpha-\beta)} \nonumber\\&&
+\frac{C}{4\rho}(2-\alpha+\beta). \nonumber
\end{eqnarray}
Consequently the model in Sect.~\ref{Model II: $a=1$} has an equation of state of the general form 
\begin{equation}
\label{M61}
p_{r}=p_{r}(\rho),
\end{equation}
which is barotropic.                                            

Another quantity of physical interest is the speed of sound $\frac{dp_r}{d\rho}$. With the help of $(\ref{M59})$ and $(\ref{M29})$, the expression for the speed of sound in terms of the radial coordinate $r$ becomes
\begin{eqnarray}
\label{M62}
\frac{dp_{r}}{d\rho}&=&\frac{2+\alpha-\beta+(2-\alpha+\beta)cr^{2}}{-10-\alpha+\beta+cr^{2}(-2-\alpha+\beta)}\nonumber\\&&
 - \frac{1}{-10-\alpha+\beta+cr^{2}(-2-\alpha+\beta)}\nonumber\\ && \times        
\left[(-1+\alpha)(1+cr^2)(2(-2 
+ cr^{2}(-1+\alpha)+\alpha)(1+\gamma^2)  \right.\nonumber\\& &\left.
+2(-1+\gamma^2)\cos(2\Omega(r))     
+6\gamma\Omega(r) \cos(2\Omega(r))  \right.\nonumber\\& &\left.
+(-3\Omega(r) \sin(2\Omega(r)) 
+\gamma(-4+3 \gamma\Omega(r))) \sin(2\Omega(r)))\right] \nonumber\\ && \times
\left[ (-1+\gamma\Omega(r)  )\cos(\Omega(r)) 
-(\Omega(r)+\gamma)\sin(\Omega(r)) \right]^{-2},
\end{eqnarray}
where $\gamma=\frac{c_{1}}{c_{2}}$ is a constant and $\Omega(r)=\sqrt{(1-\alpha)(1+Cr^2)}$ . This expression is complicated but it is interesting to note that it is possible to find an analytic expression for the speed of sound. Graphical plots can be generated for $\frac{dp_r}{d\rho}$ as we show in the next section. 

\section{Physical models\label{sec7}}
Some brief comments about the physical features of the new solutions to the Einstein-Maxwell system are made in this section. In this paper, we have presented several new models for a relativistic astrophysical star. The underlying equation was the Bessel differential equation which governs the solution of the Maxwell-Einstein system of field equations. The solutions found have matter variables which are regular and well behaved in the interior of the star. As an example we show in this section that  the exact solutions found in Sect.~\ref{Model II: $a=1$} are physically reasonable. The matter variables are plotted graphically. The software package Mathematica (Wolfram \cite{hh28}) was used for the plots with the choice of parameters $a=1$, $A=1$, $c_{1}=6.685$, $\alpha=0.59$, $\gamma=-89.87631$, $C=1$, $\beta=0.175$, $c_{2}=-0.07438$. 

For a physically realistic relativistic star the equation of state is complex and  depends
 on parameters such as the temperature, the number  fraction of a specific particle interior species and strong entropy gradients. As
 a simplifying assumption for a charged anisotropic matter distribution we assume the barotropic relationship. The expressions for the matter and geometrical variables are in terms of Bessel functions which makes it difficult to study the behaviour of the physical features. We can obtain simpler forms for these quantities by expanding them in terms of Taylor series up to order $r^2$. This makes it possible to investigate the physical behaviour for small values of $r$. We obtain the quantities:
\begin{subequations}
\label{M63}
\begin{eqnarray}
\label{M63a}
e^{2\lambda}&\approx&1+r^2,\\
\label{M63b}
e^{2\nu}&\approx& 1 + 2.90788r^{2} ,\\
\label{M63c} 
\rho&\approx& 3 - 5.2075r^2,\\
\label{M63d}
p_{r}&\approx& 5.81673 - 14.2278r^{2} , \\ 
\label{M63e}
p_{t}&\approx& 5.81673 - 14.0528r^{2},\\ 
\label{M63f} 
E^2&\approx&0.415r^2 ,\\
\label{M63g}
\sigma^2&\approx&3.735 - 16.185r^2.
\end{eqnarray}
\end{subequations}
It is clear from the above that the gravitational potentials and matter variables are finite and regular at the centre and in the spacetime region close to the centre. Note that our models in general are not asymptotically flat in the finite interior of the star; however the interior does match to the asymptotically flat vacuum exterior. To take into account the physical units and dimensional homogeneity in the plotted matter variables, we make the following scaling: $x=r^2$, $r^2=\tilde{r}^{2} /R^{2}$, $\rho=\tilde{\rho}R^2$, $p_{r}=\tilde{p_{r}}R^2$, $p_{t}=\tilde{p_{t}}R^2$, $E=\tilde{E}R^2$, $\sigma=\tilde{\sigma}R^2$, where $R$ has the dimension of length. The numerical value $R=3.75$ km has been chosen so that all the matter variables are well behaved. The quantities $ \tilde{\rho}$, $\tilde{p_{r}}$, $\tilde{p_{t}}$, $\tilde{E}$ and $\tilde{\sigma} $ are the physically relevant quantities. For example the star density is of order $3.0 \times 10^{15}$ ${\rm gcm^{-3}}$.  This is greater than the nuclear saturation density but it is in the range of
quark stars with a linear equation of state (see for example Mafa Takisa \emph{et al.} \cite{hh10}). The physical star radius is $\tilde{r}$ which is approximately $8.43$ ${\rm km}$.
Note that different choices of the parameter values may produce other physical profiles.

\begin{figure}[ht]
    \centering
		\vspace{0.1cm}
        \includegraphics[width=7.1cm,height=5.1cm]{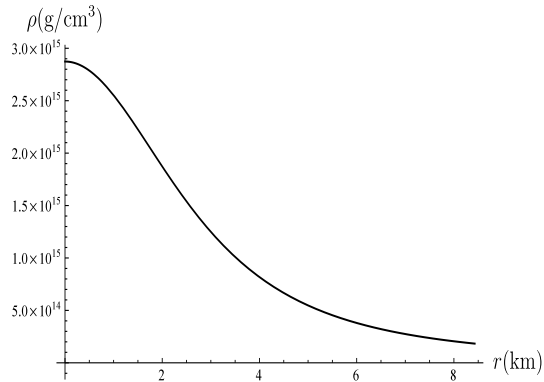} \caption{Energy density}
				\label{label:1}
\end{figure}

\begin{figure}[ht]
    \centering
		\vspace{0.1cm}
        \includegraphics[width=13.1cm,height=7.1cm]{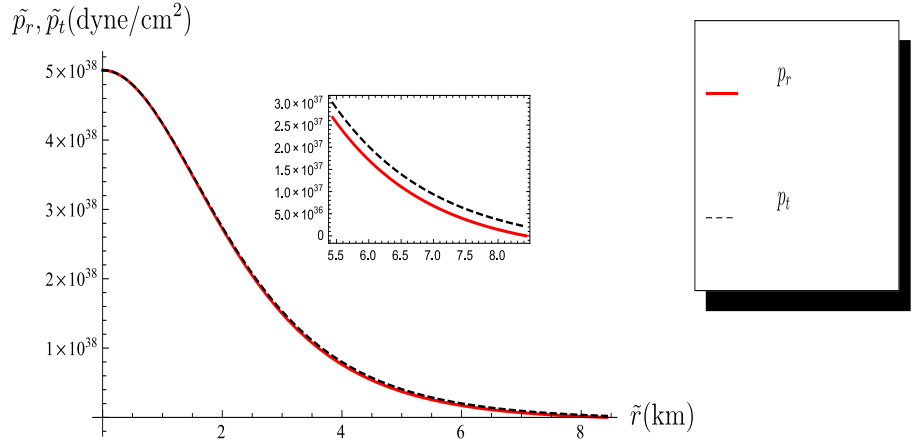} \caption{Radial and tangential pressures}
        \label{label:2}
\end{figure}

\begin{figure}[ht]
    \centering
		\vspace{0.1cm}
        \includegraphics[width=7.1cm,height=5.1cm]{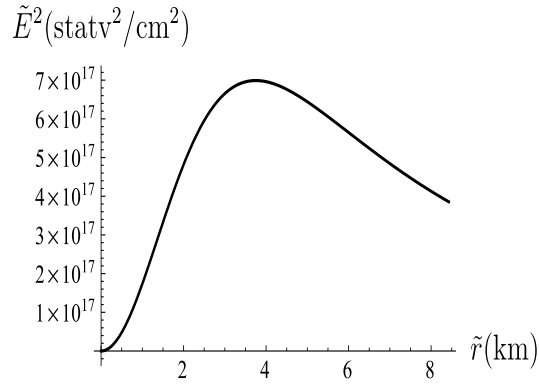} \caption{Electric field intensity}
        \label{label:3}
 \end{figure}

\begin{figure}[ht]
    \centering
		\vspace{0.1cm}
        \includegraphics[width=10.1cm,height=6.1cm]{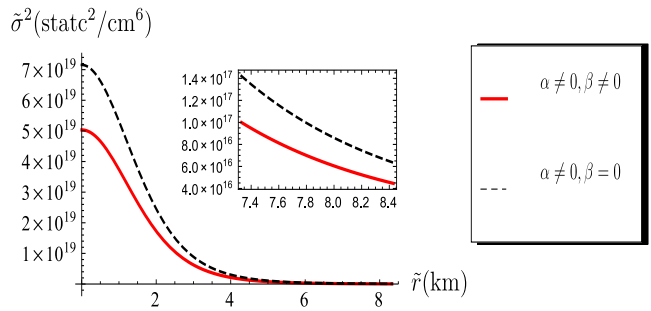} \caption{Charge density }
        \label{label:4}
\end{figure}

\begin{figure}[ht]
    \centering
		\vspace{0.1cm}
        \includegraphics[width=10.1cm,height=8.1cm]{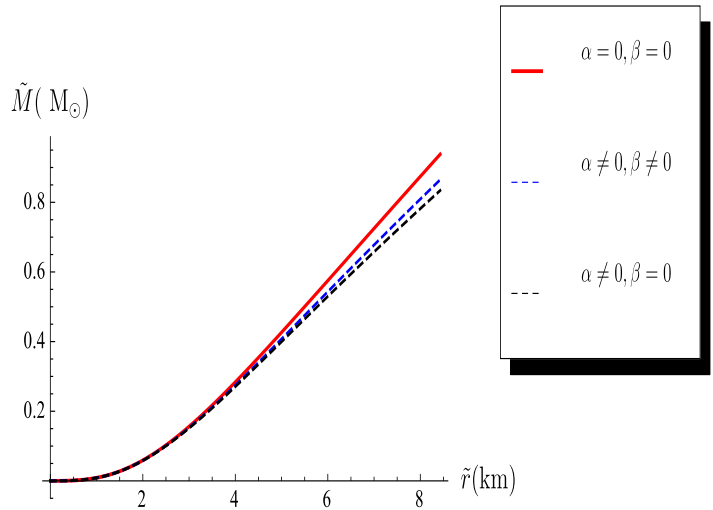} \caption{Mass function}
        \label{label:5}
 \end{figure}

\begin{figure}[ht]
    \centering
		\vspace{0.1cm}
        \includegraphics[width=10.1cm,height=7.1cm]{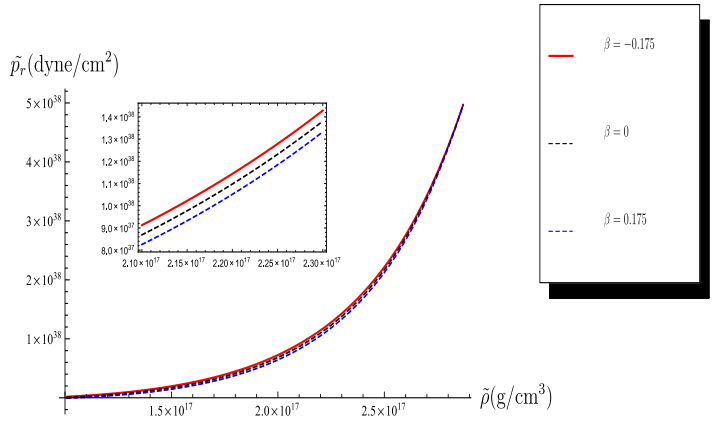} \caption{Equations of state.}
        \label{label:6}
\end{figure}

\begin{figure}[ht]
    \centering
		\vspace{0.1cm}
        \includegraphics[width=8.1cm,height=6.1cm]{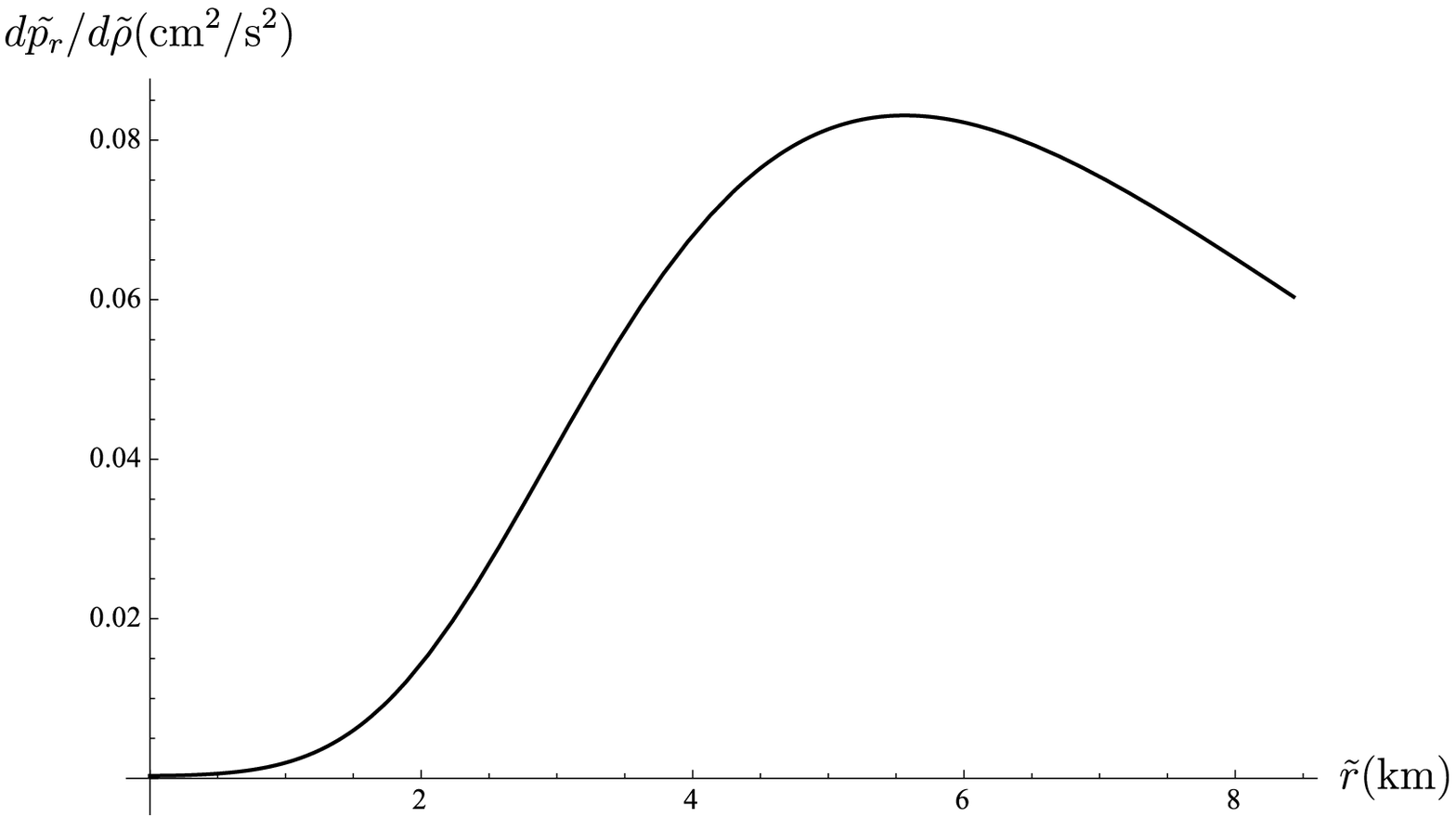} \caption{Speed of sound.}
        \label{label:7}
 \end{figure}

The following plots were generated:
\begin{itemize}
 \item Figure \ref{label:1}: Energy density.
 \item Figure \ref{label:2}: Radial and Tangential pressure.
 \item Figure \ref{label:3}: Electric field. 
 \item Figure \ref{label:4}: Charge density.
 \item Figure \ref{label:5}: Mass function.
 \item Figure \ref{label:6}: Equations of state.
 \item Figure \ref{label:7}: Speed of sound.
\end{itemize}
\noindent Figure \ref{label:1} shows that the density of energy $\tilde{\rho}$ is positive, finite and strictly decreasing. In Figure \ref{label:2} we see that both the tangential and radial pressures are positive and monotonically decreasing functions. We have introduced a zoomed box in Figure \ref{label:2} to give a better representation of the behaviour of the pressures. The zoomed box shows that while $\tilde{p_r}$ vanishes at the boundary, the tangential pressure remains positive. In Figure \ref{label:3} the electric field is positive and monotonically increasing and attains a maximum value when $\tilde{r}=R$. The evolution of the proper charge density in Figure \ref{label:4} is a decreasing function which is continuous for anisotropic and isotropic cases. The zoomed box shows the proper charge density profile near the radius. The mass function is an increasing function with increasing radius in Figure \ref{label:5}. At the radius $\tilde{r}_s=8.43 ~{\rm km}$, the ratio of mass over radius for the three cases remains in the range of $\frac{M}{\tilde{r}_s} \sim \frac{1}{10}$ to $\frac{1}{4}$ which correspond to neutron stars and ultra compact stars. We observe that the anisotropy does affect the behaviour of the mass and for these three cases. We plotted the equation of state for different parameter values in Figure \ref{label:6}. The zoomed box helps to distinguish the separation of curves for the three cases considered. We find that the parameter of anisotropy $\beta$ influences the evolution of the equation of state. In Figure \ref{label:7} the speed of sound satisfies the causality principle $0\leq\frac{dp_{r}}{d\rho}\leq 1$ and the speed of sound is less than the speed of light. The plots generated indicate that models found in this paper are physically reasonable. A detailed study of the physical features such as the luminosity and the relationship to observed astronomical objects will be carried out in future work.


\newpage

\section{ Discussion\label{sec8}}
We have comprehensively studied anisotropic and charged matter with a Finch and Skea \cite{hh13} geometry. The master equation governing the evolution of the model was derived. Several families of solution are possible to the master equation in our generalized approach. Exact solutions are possible in terms of elementary functions, Bessel functions and modified Bessel functions. When a parameter becomes an integer it is possible to represent the Bessel and modified Bessel functions in terms of elementary functions. This is demonstrated for the Bessel functions $J_{\frac{1}{2}}$, $J_{-\frac{1}{2}}$, $J_{\frac{3}{2}}$, $J_{-\frac{3}{2}}$, $J_{\frac{5}{2}}$, $J_{-\frac{5}{2}}$ and the modified Bessel functions $I_{\frac{1}{2}}$, $I_{-\frac{1}{2}}$, $I_{\frac{3}{2}}$, $I_{-\frac{3}{2}}$, $I_{\frac{5}{2}}$, $I_{-\frac{5}{2}}$. In this way an infinite family of exact solutions to the master equation can be generated in terms of elementary functions. Solutions found previously are contained in our analysis. In particular we regain the Finch and Skea \cite{hh13} solution for uncharged matter and the Hansraj and Maharaj \cite{hh14} solution in the presence of electromagnetic field. We show that the solutions found admit a barotropic equation of state so that the radial pressure can be written as a function of energy density. A graphical analysis indicates that the matter variables are well behaved and regular in the interior. In particular the speed of sound is less than the speed of light.

\section*{Acknowledgments }
DKM and PMT thank the National Research Foundation and the University of
KwaZulu-Natal for financial support.
SDM acknowledges that this work is based upon research supported by the South African Research
Chair Initiative of the Department of Science and
Technology and the National Research Foundation.

\end{document}